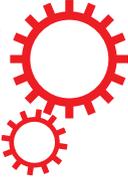

# SCIENTIFIC REPORTS

**OPEN**

# Dynamical tuning between nearly perfect reflection, absorption, and transmission of light via graphene/dielectric structures

Jacob Linder[1] & Klaus Halterman[2]

Exerting well-defined control over the reflection (*R*), absorption (*A*), and transmission (*T*) of electromagnetic waves is a key objective in quantum optics. To this end, one often utilizes hybrid structures comprised of elements with different optical properties in order to achieve features such as high *R* or high *A* for incident light. A desirable goal would be the possibility to tune between all three regimes of nearly perfect reflection, absorption, and transmission within the same device, thus swapping between the cases $R \to 1$, $A \to 1$, and $T \to 1$ dynamically. We here show that a dielectric interfaced with a graphene layer on each side allows for precisely this: by tuning only the Fermi level of graphene, all three regimes can be reached in the THz regime and below. Moreover, we show that the inclusion of cylindrical defects in the system offers a different type of control of the scattering of electromagnetic waves by means of the graphene layers.

There has recently been a surge of interest in the use of graphene as an active component in structures where the aim is to control the propagation and energy flow of electromagnetic waves[1–22]. The chief reason for this is the atomic thickness of graphene and gate-tunable Fermi level $E_F$, which allows the graphene layer to act as an effective sluice for the incident waves which depends on factors such as the electromagnetic frequency and polarization. In particular, several studies have been dedicated to how the radiation transmittance through graphene for normal incidence can be altered[23–26]. The optically high transparency of graphene combined with high electric conductivity renders this material suitable for use with ultrafast lasers[27] or as transparent electrodes[28,29]. Multilayer graphene structures, on the other hand, can be designed to feature broadband absorption with relevance for photodetectors[30,31].

Parallel with this development, there has also been a strong advancement in experimental techniques used to tailor metamaterial structures where either the permittivity tensor $\hat{\varepsilon}$, permeabitility tensor $\hat{\mu}$, or both, have vanishing components. These scenarios are known as epsilon-near-zero (ENZ), mu-near-zero (MNZ), or matched impedance zero-index metamaterials, respectively. Structures with such properties have been shown to enable unusual types of control regarding the propagation of EM waves[32–36]. Another notable example is photonic crystals which properties are known to be tunable via for instance mechanical stress[37], magnetic fields[38,39], and the incident beam intensity[40]. The combination of these two aspects, graphene and metamaterials with tunable $\hat{\varepsilon}$ and $\hat{\mu}$, are thus likely to enable a previously unattainable level of control when it comes to the scattering of light.

One of the most fundamental challenges in quantum optics is the task of designing hybrid structures that offer tunable reflection, absorption, and transmission dynamically. Our work aims at delivering a progression toward this ambitious and important goal by combining graphene with a dielectric metamaterial. By considering such a material (characterized by a permittivity $\varepsilon_1$ and permeability $\mu_1$) flanked by graphene layers on each surface, we show that one can operate the device in the THz regime and lower in three distinct regimes of high-reflectance, high-transmittance, and high-absorptance by tuning the Fermi levels in graphene. Dynamically swapping between these three regimes within a single device is highly unusual and could find use in a number of optics-related applications. We also comment in particular on limiting cases of special interest such as ENZ ($\varepsilon_1 \to 0$), and MNZ ($\mu_1 \to 0$) metamaterials. Moreover, we include the role of cylindrical rods, or "defects" embedded in a matched impedance zero-index material (MIZIM) ($\mu_1 \to 0$ and $\varepsilon_1 \to 0$), and show that defects interestingly allow for a different type of control of electromagnetic waves due to the presence of the graphene layers. The main novelty of our results compared to previous literature is that our proposed device, including only graphene and a dielectric layer, can be used to tune between the three distinct regimes of nearly perfect reflection, absorption, and transmission of electromagnetic waves, as opposed to operating in merely one of these regimes. Due to the simplicity of the structure, our results should be feasible to check experimentally and may find use in applications requiring a strong element of control of the reflection, absorption, and transmission of electromagnetic waves.

[1]Department of Physics, NTNU, Norwegian University of Science and Technology, N-7491 Trondheim, Norway. [2]Michelson Lab, Physics Division, Naval Air Warfare Center, China Lake, California 93555, USA. Correspondence and requests for materials should be addressed to J.L. (email: jacob.linder@ntnu.no)

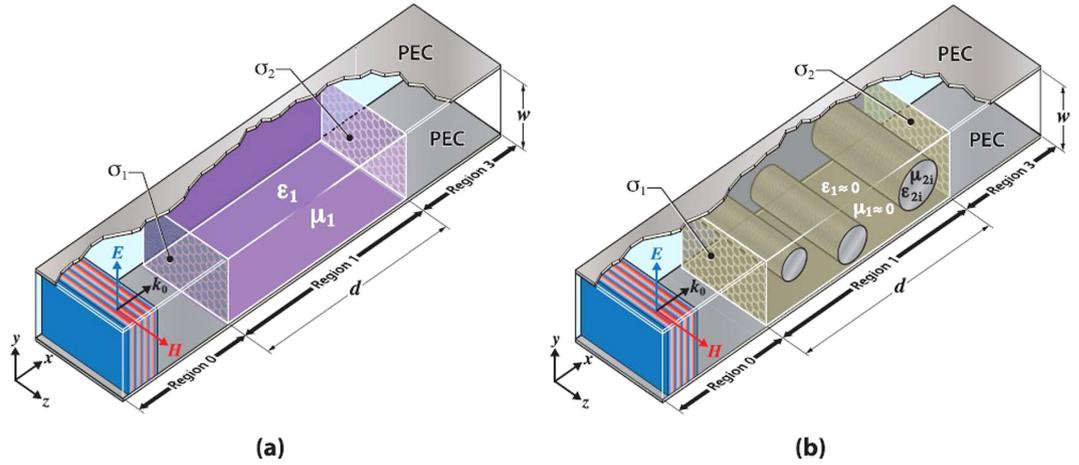

**Figure 1. Proposed experimental setup.** Schematic of the double graphene metamaterial configurations investigated in this paper. The incident wave in region 0 has propagation constant $k_0$ directed along $x$, and for the TEM modes considered in this paper, **E** is directed along $y$ and **H** is directed along $z$. Each graphene sheet separated by a distance $d$ has conductivity $\sigma_1$ and $\sigma_2$, as shown. The system is grounded by two perfect electric conductors (PECs) that are a distance $w$ apart. (**a**): The uniform media in region 1 has permittivity $\varepsilon_1$ and permeability $\mu_1$. (**b**) Region 1 is composed of matched impedance zero index media (MIZIM) with $\varepsilon_1 \to 0$ and $\mu_1 \to 0$.

## Dielectric with graphene boundaries

We consider first the system shown in Fig. 1a. For an incident TEM mode, one may write down the general expressions for the magnetic (**H**) and electric (**E**) fields in each of the regions 0–3. The graphene layers are taken into account through the boundary condition $\mathbf{n} \times (\mathbf{H}_{i+1} - \mathbf{H}_i) = \sigma_j \mathbf{E}_\perp$ where **n** is a unit vector pointing from region $i$ to $(i+1)$, and $\mathbf{E}_\perp$ is the component of the electric field perpendicular to the interface. The correctness of modeling graphene as a surface conductivity sheet as opposed to using a slab model to treat the optical properties of graphene has recently been discussed and confirmed in ref. 41. The complex ac conductivities of the graphene layers $\sigma_j$ ($j = 1, 2$) can vary depending on their respective gate voltages. Additional details of the calculations are given in the Methods section. We present here the final result for the reflection and transmission coefficients:

$$r = \frac{s \cos l (q_1 + q_2) + i \sin l (1 + q_2 - q_1 - q_1 q_2 - s^2)}{s \cos l (2 + q_1 + q_2) - i \sin l (1 + q_1 + q_2 + q_1 q_2 + s^2)}, \quad (1)$$

$$t = \frac{2s}{s \cos l (2 + q_1 + q_2) - i \sin l (1 + q_1 + q_2 + q_1 q_2 + s^2)}. \quad (2)$$

From Eqs (1) and (2), one obtains the reflection ($R$), transmission ($T$), and absorption ($A$) probabilities according to $R = |r|^2$, $T = |t|^2$, and $A = 1 - |r|^2 - |t|^2 \equiv |a|^2$. Above, $l = k_1 d$ where $d$ is the length of the dielectric region, $q_i = \sigma_i/(c\varepsilon_0)$ is proportional to the conductivities at the interfaces, $s = \sqrt{\varepsilon_1 \mu_0/(\varepsilon_0 \mu_1)}$, $k_0 = \omega \sqrt{\varepsilon_0 \mu_0}$, and $k_1 = \omega \sqrt{\varepsilon_1 \mu_1}$, where $\omega$ is the frequency.

Before turning to our main results, some observations can be made regarding Eqs (1) and (2). The transmission coefficient $t$ is invariant when interchanging the graphene layers, $q_1 \leftrightarrow q_2$, whereas the reflection coefficient is not, except in certain limiting cases, such as the limit of electrically small layers $d/\lambda \ll 1$, leading to $r \simeq \frac{q_1 + q_2}{2 + q_1 + q_2}$, and $t \simeq \frac{2}{2 + q_1 + q_2}$. In the discussion that follows, and in the numerical calculations used in this paper, we will consider the case where $\text{Im}(q_i) \ll \text{Re}(q_i)$, allowing us to consider limits such as $q_i \simeq 1$ which will turn out to be of special interest. To ensure the validity of this, we will use parameters so that $\omega t_{\text{rel}} \ll 1$ in the plots (see below for definition of $t_{\text{rel}}$), which provides precisely $\text{Im}(q_i) \ll \text{Re}(q_i)$.

The ENZ and MNZ cases are of particular interest, which correspond to $s \to 0$ and $s \to \infty$, respectively. In the MNZ ($s \to \infty$) case, we find $t = \frac{2}{2 - i\varepsilon_1 k_0 d/\varepsilon_0 + q_1 + q_2}$ and $r = 1 - \frac{2}{2 - i\varepsilon_1 k_0 d/\varepsilon_0 + q_1 + q_2}$, so that for electrically thick layers, $d/\lambda \gg 1$, or when $q_{1,2} \gg 1$, there is no transmission and perfect reflection. For the ENZ ($s \to 0$) scenario, we find

$$r = \frac{k_0 d(q_1 - 1)(q_2 + 1) + i(q_1 + q_2)}{k_0 d(q_1 + 1)(q_2 + 1) + i(2 + q_1 + q_2)}, \quad (3)$$

and

$$t = \frac{2}{2 + q_1 + q_2 - ik_0 d(q_1 + 1)(q_2 + 1)}, \quad (4)$$

where $\mu_1 = \mu_0$. Therefore if the conductivity in the first graphene sheet is large, $q_1 \gg 1$, we get total reflection. Increasing $q_2$ to large values, again results in no transmission, however the reflection is tunable via $q_1$: $r = 1 - \frac{2k_0 d}{i + k_0 d(1 + q_1)}$. In the case for thick ENZ films $d/\lambda \gg 1$, we get $r \simeq \frac{q_1 - 1}{q_1 + 1}$, and $t = 0$. Here, the relevant variations in $R$ or $A$ are due to $\sigma_1$. Indeed complete absorption can occur when $q_1 = 1$, or equivalently when $\sigma_1 = c\varepsilon_0$. This result is consistent with an incident beam normally incident on a graphene-based ENZ structure[42]. Therefore, when a metamaterial with extremely small values for its permittivity or permeability is sandwiched between two graphene sheets, it follows that one can use the corresponding graphene conductivities $q_i$ to tune between strong reflection, transmission, or absorption.

Arguably the most interesting regime somewhat surprisingly turns out to be the most "conventional" one, where the dielectric has a relative permittivity and permeability close to 1, or impedance matched with $\sqrt{\mu_1/\varepsilon_1} = \sqrt{\mu_0/\varepsilon_0}$, so that $s \simeq 1$. It is in this case we discover that one may dynamically tune between the regimes of nearly perfect $R$, $A$, and $T$, as we now proceed to show. For a given structure, the value of $d$ and $s$ are locked and likely to be very difficult to tune dynamically. We thus fix the values of these quantities in what follows. To illustrate the versatility offered by the graphene layers in an analytically clear way, we first consider the case where $l = k_1 d = \pi/2 + 2n\pi$, $n = 0, \pm 1, \pm 2, \ldots$. Note that this means that we are not limiting the consideration to only one frequency $\omega$: the transition between the $R$, $A$, and $T$ regimes can be obtained for a range of frequencies due to the periodic nature of $l$. We provide results for the more general case where $l$ is unrestricted in the Methods section, where the system can also transit between nearly perfect $R$, $A$, and $T$. We obtain with $l = k_1 d = \pi/2 + 2n\pi$ that:

$$r = \frac{q_1 q_2 + s^2 + q_1 - q_2 - 1}{1 + q_1 + q_2 + q_1 q_2 + s^2},$$
$$t = \frac{2is}{1 + q_1 + q_2 + q_1 q_2 + s^2}. \quad (5)$$

The choice $s \simeq 1$ should be feasible experimentally in the THz regime (we discuss this in more detail later and for now only comment that using experimentally realized low-loss doped polymers[43] could be a route to this end) and we focus on this for concreteness in what follows. Moreover, since we consider frequencies in the THz region and lower, where the intraband part to the conductivity dominates the interband part, we take each graphene sheet to have a conductivity, $\sigma_i = 2ie^2 k_B \widetilde{T}/(\pi \hbar^2)(\omega + i/t_{rel})^{-1} \ln[2 \cosh(E_{Fi}/(2k_B\widetilde{T}))]$. Here $E_{Fi}$ is the Fermi energy of a given graphene sheet ($i = 1, 2$), $\omega$ is the frequency of the EM wave, $\widetilde{T}$ is the temperature, and $t_{rel}$ is the relaxation time (assumed to be the same for both graphene layers). Thus when $E_F \gg k_B \widetilde{T}, \hbar\omega$, we can write simply,

$$\sigma = ie^2 E_F / [\pi \hbar^2(\omega + i/t_{rel})]. \quad (6)$$

From Eq. (5), one then finds that the following relations hold by altering only the Fermi levels of the graphene layers:

- **Perfect transmission:** $|t| \to 1$ when $q_1 \ll 1$ and $q_2 \ll 1$.
- **Perfect reflection:** $|r| \to 1$ when $q_1 \gg 1$.
- **Perfect absorption:** $t \to 0$, $r \to 0$ when $q_2 \gg 1$ and $q_1 \simeq 1$.

One may also write down the conditions for the same three scenarios without limiting $l$, in which case the analytical expressions are more comprehensive (see Methods section). It is remarkable that one can dynamically tune between all three regimes in a single structure and for a range of frequencies. This is shown graphically in Fig. 2, where a sphere in the parameter space spanned by the magnitudes of ($r$, $t$, $a$) is shown. The two paths on the sphere depict how to acheive the desired admixtures of $R$, $T$, and $A$, by simply tuning the relative Fermi levels in the graphene sheets in the graphene/dielectric/graphene system. The first path (green curve) begins with the system in a highly reflective state as both graphene layers are gate tuned to equal levels of 800 meV. As the Fermi level in each graphene sheet is reduced by equal amounts, the structure then permits substantial transmission of the incident beam. Finally, by tuning $E_{F1}$ and $E_{F2}$ to be considerably different, we end up with a system that is a near perfect absorber. In panel (**a**) the reflectance is shown while the Fermi energies in the graphene layers are each varied. A wide range of energies are considered, from 0 meV, up to 800 meV. In this plot, it is evident that $R$ can evolve from a state of no reflection to one of nearly perfect reflection when just increasing $E_{F1}$ and keeping $E_{F2}$ fixed. The conductivity of the first graphene layer nearest the incident beam clearly plays the bigger role here, as increases in $E_{F1}$ cause $\sigma_1$ to become large (both real and imaginary parts increasing nearly linearly with $E_{F1}$). The corresponding impedance mismatch between the two regions results in the observed enhanced reflectivity. Next, in Fig. 2b, we illustrate how the transmission can be manipulated through changes in the Fermi energies. Since the incident wave must pass through both graphene sheets effectively for there to be high transmission, it is evident that the Fermi levels must be very small in both sheets so that $\sigma_1, \sigma_2 \to 0$, or equivalently when $q_1 \ll 1$ and $q_2 \ll 1$. Next, for the structure to be a perfect absorber, it is of course necessary for there to not be any reflection or transmission of the incident wave. Figure 2c illustrates the conditions on the gate tunable Fermi energies for optimal absorption. We know that in order for reflection to be minimized, the incident beam should efficiently couple to the first graphene sheet, which occurs when $q_1 \simeq 1$. Then, by raising $E_{F2}$ (so that $q_2 \gg 1$), the transmission of waves can then be restricted from entering the vacuum region 3 by the second graphene layer.

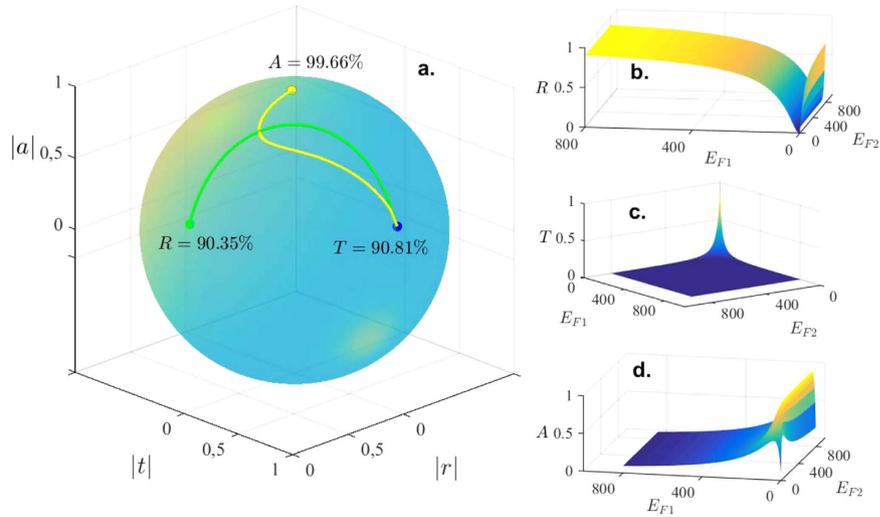

**Figure 2. Tunable transmission, reflection, and absorption in a graphene/dielectric/graphene layer.** The parameters are chosen as $s = 1.01$, $t_{rel} = 1.1 \times 10^{-12}$ s, $\omega = 0.1$ THz, and $d = \pi/(2k_1)$. (**a**) Demonstration of the transition between the three regimes. Starting at $E_{F1} = E_{F2} = 800$ meV, the device is in the high-reflectance operational mode ($R = 90.35\%$). Lowering the two Fermi levels to $E_{F1} = E_{F2} = 1$ meV creates a path (green line) to the point with high transmittance ($T = 90.81\%$). Finally, increasing again the Fermi levels to $E_{F1} = 20.31$ meV and $E_{F2} = 800$ meV renders the device a nearly perfect absorber via the yellow path ($A = 99.66\%$). (**b**) The reflectance $R$ as a function of the Fermi levels (measured in meV) in the two graphene layers. The transmittance $T$ and absorptance $A$ are shown in (**c**) and (**d**) respectively. The temperature satisfies $k_B \widetilde{T} \ll E_F$.

The distribution of energy can be accounted for by using Poynting's theorem for time harmonic fields[44], which for the $s \simeq 1$ regime is simply: $2i\omega \mathcal{U} + \partial S_x/\partial x = 0$. The quantity $\mathcal{U}$ is defined in terms of the time averaged electric and magnetic energy densities $u_E$ and $u_H$, respectively: $\mathcal{U} \equiv u_E - u_H$, and the $x$-component of the time averaged complex Poynting vector $S_x$ is given by $S_x = 1/2 E_y H_z^*$. We focus on the central region so that the real part of $S_x$ represents the energy flow within that segment, and the time averaged energy densities $u_E$ and $u_H$ correspond to the respective electric and magnetic field contributions, with $u_E = 1/4\varepsilon_1 |E_y|^2$, and $u_H = 1/4\mu_1 |H_z|^2$. The net energy density is due to the energies of the forward and backward waves, whereas $\Re\{S_x\}$ can be used to give the energy flow direction arising from the difference between two countermoving waves. For $s \simeq 1$, we find, $\Re\{S_x\} \simeq 2\sqrt{\mu_0/\varepsilon_0}(1 + \Re\{q_2\})/|2 + q_1 + q_2 + q_1 q_2|^2$, so that the energy flow in region 1 is always directed towards the positive $x$ direction (towards the second graphene sheet), with no net flow across the graphene sheets. This is consistent with the expression for $\mathcal{U}$, which is purely real:

$$\mathcal{U} = \frac{-2\mu_0 [(|q_2|^2 + 2\Re\{q_2\})\sin(\pi x/d) - 2\Im\{q_2\} \cos(\pi x/d)]}{|2 + q_1 + q_2 + q_1 q_2|^2}, \qquad (7)$$

demonstrating from the conservation equation, the vanishing of the real part of $\partial S_x/\partial x$.

### Cylindrical defects

If the region 1 is a MIZIM, whereby $\varepsilon_1 \to 0$ and $\mu_1 \to 0$, the **E** and **H** fields become spatially uniform: $E_{y,1} \simeq \sqrt{\mu_0/\varepsilon_0}\, t$, and $H_{z,1} \simeq (1 + q_2)t$, where,

$$r = \frac{q_1 + q_2}{2 + q_1 + q_2},$$

$$t = \frac{2}{2 + q_1 + q_2}. \qquad (8)$$

Therefore $r$ and $t$ are each independent of the material thickness, and equivalent to an electrically small region discussed above. We consider here a frequency dispersive permittivity and permeability in region 1, with $\omega \to \omega_p$, where $\omega_p$ is a characteristic frequency of the Drude responses: $\varepsilon_1 = \varepsilon_0(1 - \omega_p^2/[\omega(\omega + i\Gamma)])$ and $\mu_1 = \mu_0(1 - \omega_p^2/[\omega(\omega + i\Gamma)])$. For simplicity, we set $\Gamma = 0$. It is evident from Eq. (8) that if the graphene sheets are initially set so that $q_{1,2} = 0$, the system has complete transmission. By increasing the gate voltage so that the conductivities increase to levels corresponding to $q_{1,2} \gg 1$, we get $t \to 0$ and $r \to 1$.

We now consider the insertion of dielectric rods, or "defects" into the central metamaterial MIZIM region, where a different type of control over the transmission properties of the TEM mode can be achieved. Considering the system shown in Fig. 1b, where region 1 is a MIZIM, we obtain the following transmission and reflection coefficients (details are given in the Methods section):

$$t = \frac{1}{1 + (q_1 + q_2)/2 - i(1 + q_1)(1 + q_2)\Sigma}, \quad (9)$$

$$r = \frac{(q_1 + q_2)/2 - i(q_1 - 1)(q_2 + 1)\Sigma}{1 + (q_1 + q_2)/2 - i(1 + q_1)(1 + q_2)\Sigma} \quad (10)$$

The physical effect of the cylindrical defects is captured in the quantity $\Sigma$, which is purely real and can take any sign, defined as ref. 32:

$$\Sigma = \frac{\pi}{w} \sum_{m=1}^{N} R_m \sqrt{\frac{\mu_{2m}/\mu_0}{\varepsilon_{2m}/\varepsilon_0}} \frac{J_1(k_{2m}R_m)}{J_0(k_{2m}R_m)}. \quad (11)$$

Here, $w$ is the width of the sample, and the summation is taken over the number of defects. A given defect $m$ has radius $R_m$, and wavenumber $k_{2m} = \omega\sqrt{\varepsilon_{2m}\mu_{2m}}$, while $J_0$ and $J_1$ are Bessel functions of the first kind of order 0 and 1, respectively.

For a given frequency, the value of $\Sigma$ is assumed fixed in what follows. If the quantity $k_{2m}R_m$ corresponds to a zero of the $J_0$ or $J_1$, then Eq. (11) shows that $\Sigma$ diverges or vanishes, thus illustrating the extreme range of values it can possess. It is evident from Eq. (9) that the transmission coefficient is again invariant under exchange of the two graphene conductivities, $q_1 \leftrightarrow q_2$. The reflection coefficient is, on the other hand, not invariant under the exchange $q_1 \leftrightarrow q_2$ except when $\Sigma = 0$ corresponding to an absence of defects, or if defects are present, each one satisfies $J_1(k_{2m}R_m) = 0$. This is consistent with Eq. (8) which is equivalent to Eq. (9) when $\Sigma = 0$.

From Eq. (9), we see that the role of $\Sigma$ with regard to transmission always is to reduce $t$ as long as $q_{1,2} \neq 0$. Interestingly, we can swap between perfect reflection and perfect absorption for a given $\Sigma$. To see this, consider the case $\Sigma \gg 1$, which occurs when *any* defect has a radius and material parameters that satisfy $J_0(k_{2m}R_m) = 0$. In that case, $t \to 0$, $r \to \frac{q_1 - 1}{q_1 + 1}$, so that there is vanishing reflectivity if $q_1$ is tuned such that $q_1 \simeq 1$. This result is equivalent to the ENZ case above with a thick interlayer. In this case, the absorption is expressed as $A = \frac{4q_1}{(1 + q_1)^2}$, which clearly shows perfect absorption when $q_1 = 1$.

The relevant parameters $R$, $T$, and $A$ are shown in the left column of Fig. 3. Note that the observed enhanced absorption in (c) arises from the interference of waves, since this result holds in the absence of losses in region 1. On the other hand, we get $t = 0$ and $r = 1$ in the limit $q_1 \gg 1$. It follows then that when the defects are designed so that $\Sigma \gg 1$, the first graphene sheet described by $q_1$ completely controls the reflection and absorption of EM waves.

If the cylinders are now chosen so that $\Sigma = 0$, then the $r$ and $t$ coefficients revert to the expressions shown in Eq. (8), where $r$ and $t$ are directly tunable via manipulating only $q_1$ and $q_2$. In the right column of Fig. 3, we see that by first tuning the gate voltages so that $q_{1,2} = 0$, there is total transmission. Increasing the Fermi levels in one or both graphene sheets then increases the reflections of waves until $q_{1,2} \gg 1$, at which point nearly complete reflection occurs. As the right column of Fig. 3 further shows, for low values of the Fermi levels, we get nearly perfect transmission, and zero reflection, and absorption. The absorption is seen to never exceed 0.5 over the whole range of $E_{F_1}$ and $E_{F_2}$. Indeed, when either $q_1$ or $q_2$ is small, we find $A = \frac{4q_i}{(2 + q_i)^2}$ (assuming the imaginary part to $q_i$ is small). This quantity has a maximum when $q_i = 2$, which gives $A = 0.5$.

We can moreover control the reflectivity of the system in situations when the rods are not excited at resonance. In particular, if we relax the restriction that $\Sigma \gg 1$, setting the numerator of Eq. (10) equal to zero gives the relationship between $q_1$ and $q_2$ that results in no reflections from the first graphene sheet. For example we get $r = 0$ if $q_2 = 0$, and $q_1 = \frac{2\Sigma}{i + 2\Sigma}$.

## Discussion

Other works have shown high-absorption of normal incidence light by using *e.g.* three graphene layers separated by a dielectric spacer and an ENZ metamaterial[45], although such a structure is considerably more complex than the one proposed in this work. Regarding material choice for the dielectric in Fig. 1(a), elastomeric polymers are often used for THz applications[46–48], albeit these have moderate loss in this frequency range[49,50]. However, a recent work reported a technique for producing an effective low-loss dielectric media for THz waves by combining elastomeric polydimethylsiloxane with dopants $Al_2O_3$ and polytetrafluoroethylene[43]. At 0.7 THz, the extracted material properties were of order $\varepsilon_r \simeq 2$ and a loss tangent of order $\tan \delta \simeq 5 \times 10^{-2}$, thus being close to the prescribed dielectric response used in the present manuscript. In fact, using such a doped polymer would correspond to setting $s \simeq 2$ in our previous treatment in which case the conclusions for perfect reflection, absorption, and transmission obtained as the graphene Fermi level is tuned are unchanged, the only exception being that $|t| \to 0.8$ instead of $|t| \to 1$ in the transmission regime.

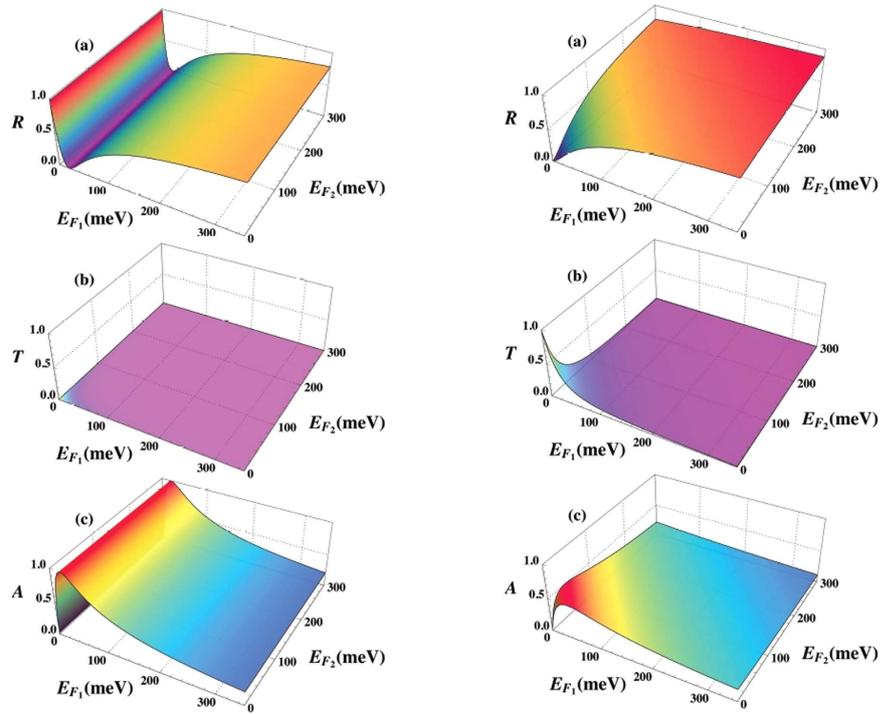

**Figure 3. Tunable transmission, reflection, and absorption in a graphene/MIZIM/graphene layer with cylindrical defects.** *Left column:* A single cylindrical dielectric rod of radius $R \approx 3.6$ mm is considered with permittivity of $\varepsilon_2 = 4\varepsilon_0$, which for the operational frequency of $\omega = 0.1$ THz yields $\Sigma \gg 1$. We show (**a**) the reflectance $R$ (**b**) the transmittance $T$, and (**c**) the absorptance $A$ as a function of the Fermi levels (measured in meV) in the two graphene layers. *Right column:* A single cylindrical dielectric rod of radius $R \approx 5.7$ mm is considered with permittivity of $\varepsilon_2 = 4\varepsilon_0$, which for the operational frequency of $\omega = 0.1$ THz yields $\Sigma \simeq 0$. We show **a.** the reflectance $R$ (**b**) the transmittance $T$, and (**c**) the absorptance $A$ as a function of the Fermi levels (measured in meV) in the two graphene layers.

Regarding the tunable Fermi level of graphene, one should acknowledge the experimental difficulty in minimizing fluctuations $\delta E_F$ in the local Fermi level which typically are a few tens of meV. However, recent work using both single-layer[51] and bilayer graphene[52] have shown that these fluctuations can be strongly suppressed, in which case our results for $E_F \simeq 1$ meV are of relevance. For the smallest Fermi levels $E_F$ considered in our work, the temperature needs to be less than 10 K in order to satisfy the requirement $E_F \gg k_B \tilde{T}$ used in the expression for the conductivity.

If the central region is a MIZIM [Fig. 1(b)], there are a number of experimental possibilities for creating a system that has an effective EM response corresponding to $\varepsilon \to 0$ and $\mu \to 0$ in the microwave and infrared regimes[53,54]. One possible approach involves the use of vertically stacked silicon rods that are separated by silicon dioxide to create an isotropic, low loss, resonant all-dielectric metamaterial[54]. Alternatively, arrays of dielectric rods comprising a dielectric photonic crystal can be designed with conventional dielectric materials to possess a Dirac-like dispersion that yields a MIZIM response near the frequency coinciding with the Dirac point[53]. Moreover, if the dielectric rods are embedded in an anisotropic high permittivity background, the MIZIM response can be accessible to all polarizations. Although designing a metamaterial within effective medium theory to possess an effectively simultaneous zero permittivity and permeability in the THz regime can pose challenges, other experimental implementations may also be possible with multilayer dielectrics or rod lattices[55] built with polaritonic components[56,57].

## Methods

We here provide details for the solution of the Maxwell equations and the scattering coefficients.

*Dielectric case*: we here provide some technical details regarding the calculation of the reflectance, transmittance, and absorptance of an air/graphene/dielectric/graphene/air region. For an incident TEM mode, we use the following fields (omitting the vector character, as $H$ is always in the $\hat{z}$ direction and $E$ is always in the $\hat{y}$-direction):

$$H_0 = h_0[e^{ik_0(x+d/2)} + re^{-ik_0(x+d/2)}], \qquad (12a)$$

$$E_0 = \frac{h_0}{c\varepsilon_0}\left[e^{ik_0(x+d/2)} - re^{-ik_0(x+d/2)}\right], \qquad (12b)$$

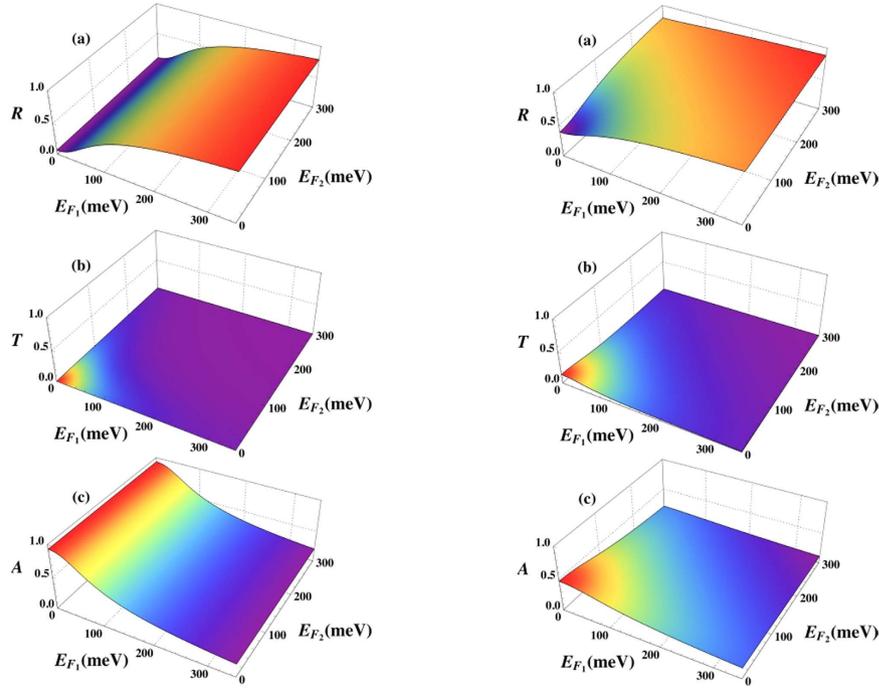

**Figure 4. The role of temperature for tunable transmission, reflection, and absorption in a graphene/MIZIM/graphene layer with cylindrical defects.** Comparison with Fig. 3 (where $E_F \gg k_B \widetilde{T}$ was assumed) using now instead room-temperature ($\widetilde{T} = 290$ K) where the full temperature-dependent expression for $\sigma_i$ is used. *Left column:* A single cylindrical dielectric rod of radius $R \approx 3.6$ mm is considered with permittivity of $\varepsilon_2 = 4\varepsilon_0$, which for the operational frequency of $\omega = 0.1$ THz yields $\Sigma \gg 1$. We show (**a**) the reflectance $R$ (**b**) the transmittance $T$, and c. the absorptance $A$ as a function of the Fermi levels (measured in meV) in the two graphene layers. *Right column:* A single cylindrical dielectric rod of radius $R \approx 5.7$ mm is considered with permittivity of $\varepsilon_2 = 4\varepsilon_0$, which for the operational frequency of $\omega = 0.1$ THz yields $\Sigma \simeq 0$. We show a. the reflectance $R$ (**b**) the transmittance $T$, and (**c**) the absorptance $A$ as a function of the Fermi levels (measured in meV) in the two graphene layers.

$$H_1 = h_0[\tilde{a}e^{ik_1 x} + \tilde{b}e^{-ik_1 x}], \tag{12c}$$

$$E_1 = h_0 \sqrt{\frac{\mu_1}{\varepsilon_1}} \left[\tilde{a}e^{ik_1 x} - \tilde{b}e^{-ik_1 x}\right], \tag{12d}$$

$$H_3 = h_0 t e^{ik_0(x-d/2)}, \tag{12e}$$

$$E_3 = \frac{h_0}{c\varepsilon_0} t e^{ik_0(x-d/2)}, \tag{12f}$$

where $k_0$ is the free space wavenumber: $k_0 = \omega/c = \omega\sqrt{\varepsilon_0 \mu_0}$, $h_0$ is the magnitude of the **H**-field, and the harmonic time-dependence $e^{i\omega t}$ has been suppressed. In all calculations, we normalized the fields so that $h_0 = 1$. The boundary conditions, obtained by using the general form

$$\boldsymbol{n} \times (\boldsymbol{H}_1 - \boldsymbol{H}_0) = \sigma \boldsymbol{E}_\perp, \tag{13}$$

where $\boldsymbol{n}$ is the unit vector pointing from region 0 to 1 and $\perp$ means the component perpendicular to the interface, become the following in our case. At $x = -d/2$:

$$H_0 - H_1 = \sigma_1 E_0 \text{ and } E_0 = E_1. \tag{14}$$

At $x = d/2$:

$$H_1 - H_3 = \sigma_2 E_3 \text{ and } E_1 = E_3. \tag{15}$$

For a dielectric material with graphene at both the interfaces, using permittivity and permeability $\varepsilon_1$ and $\mu_1$ for the dielectric region, we obtain the results in Eqs (1) and (2). The coefficients $\tilde{a}$ and $\tilde{b}$ are found similarly:

$$\tilde{a} = \frac{e^{-il/2}}{2}(1 + q_2 + s)t \tag{16a}$$

$$\tilde{b} = \frac{e^{il/2}}{2}(1 + q_2 - s)t, \tag{16b}$$

where $t$ is given in Eq. (2)

When $l$ is arbitrary, we find (using $s \simeq 1$ for concreteness) that $q_1$ and $q_2$ need to satisfy the following relations.

Perfect transmission $|t| \to 1$:

$$|\cos l(2 + q_1 + q_2) - i \sin l(2 + q_1 + q_2 + q_1 q_2)| = 2. \tag{17}$$

Perfect reflection: $|r| \to 1$:

$$\begin{aligned} &\cos l(q_1 + q_2) + i \sin l(1 + q_1 + q_2 - q_1 q_2 - s^2) \\ &= \pm[\cos l(2 + q_1 + q_2) - i \sin l(2 + q_1 + q_2 + q_1 q_2)]. \end{aligned} \tag{18}$$

Perfect absorption: $t \to 0, r \to 0$:

$$\begin{aligned} &\cos(q_1 + q_2) + i \sin l(q_2 - q_1 - q_1 q_2) = 0 \text{ and} \\ &|\cos l(2 + q_1 + q_2) - i \sin l(2 + q_1 + q_2 + q_1 q_2)| \gg 1. \end{aligned} \tag{19}$$

For the ENZ case, the electric field varies linearly in the $x$ coordinate:

$$E_1 \simeq \frac{2\sqrt{\frac{\mu_0}{\varepsilon_0}}(1 + ik_0(1 + q_2)(x - d/2))}{2 + q_1 + q_2 - ik_0 d(1 + q_1)(1 + q_2)}, \tag{20}$$

while the magnetic field is uniform throughout the entire ENZ region:

$$H_1 \simeq \frac{2(1 + q_2)}{2 + q_1 + q_2 - ik_0 d(1 + q_1)(1 + q_2)}, \tag{21}$$

*Cylindrical defects case*: when cylindrical defects are included and the dielectric material is now a MIZIM (see Fig. 1b), we proceed along the lines of ref. 32. In the vacuum regions 0 and 3, the electric and magnetic fields are written as before in Eqs (12a)-(12b), and Eqs (12e)-(12f). The Maxwell equation for a given region $j$, $\boldsymbol{E}_j = i/(\omega\varepsilon_j)\boldsymbol{\nabla} \times \boldsymbol{H}_j$ reveals that for the MIZIM region where $\varepsilon_1 \to 0$, we must have $\boldsymbol{\nabla} \times \boldsymbol{H}_1$ vanish so that the electric field remains finite. Therefore in region 1, we write simply $H_1 = h_1$, where $h_1$ is a constant ($\boldsymbol{H}_1$ is in the $\hat{\boldsymbol{z}}$ direction). Next, utilizing the boundary conditions Eqs (14) and (15), we arrive at the following relationships between the EM field coefficients:

$$h_0(1 + r) - h_1 = q_1 h_0(1 - r) \tag{22}$$

$$h_1 - h_0 t = q_2 h_0 t. \tag{23}$$

Combining these equations gives a simple expression involving $r$ and $t$:

$$(1 + r) - t(q_2 + 1) = q_1(1 - r) \tag{24}$$

Since within the MIZIM region we have a vanishing permeability $\mu_1$, the Maxwell equation $\boldsymbol{H}_j = -i/(\omega\mu_j)\boldsymbol{\nabla} \times \boldsymbol{E}_j$ shows that $\boldsymbol{\nabla} \times \boldsymbol{E}_1$ must also vanish, for physical reasons that are similar to those requiring the vanishing of the curl of $\boldsymbol{H}_1$ above. If we apply Stoke's theorem to the outer boundary $\partial C$ of the MIZIM structure, which includes the PEC layers and graphene sheets, as well as the circular boundaries $\partial C_{2m}$ of each of the defects, the contributing paths in the line integral lead to the following expression involving the corresponding $\boldsymbol{E}$ field:

$$\oint_{\partial C} \boldsymbol{E} \cdot \boldsymbol{dl} + \sum_{m=1}^{N} \oint_{\partial C_{2m}} \boldsymbol{E}_{2m} \cdot \boldsymbol{dl} = 0. \tag{25}$$

The $\boldsymbol{E}$ fields for the $\partial C$ segments has been given above in Eqs (12b–12f). The electric field within each cylindrical defect $E_{2m}$ involves a sum of Bessel functions of the first kind[32],

$$\boldsymbol{E}_{2m} = ih_1 \sqrt{\frac{\mu_{2m}}{\varepsilon_{2m}}} \frac{J_1(k_{2m} r_m)}{J_0(k_{2m} R_m)} \hat{\boldsymbol{\phi}}_m, \tag{26}$$

where $r_m$ is the radial coordinate relative to the center of each defect of radius $R_m$, and $\hat{\boldsymbol{\phi}}_m$ is the azimuthal unit vector for the $m$th cylinder. Inserting the electric fields into the line integral in Eq. (25) yields,

$$1 - r - t + 2it(1 + q_2)\Sigma = 0, \qquad (27)$$

where we have used Eq. (23), and $\Sigma$ is given in Eq. (11). Next we use Eq. (24) and Eq. (27) to solve for either $r$ or $t$, yielding the results found in Eqs (9) and (10).

Finally, we also provide plots that show the effect of relaxing the criterion $E_F \gg k_B T$, i.e. the role of temperature. This is shown in Fig. 4 where we have considered the same parameter set as in Fig. 3, but set the temperature to $\tilde{T} = 290$ K.

## References

1. S. H. Lee *et al.* Switching terahertz waves with gate-controlled active graphene metamaterials. *Nat. Mater.* **11,** 936 (2012).
2. C. S. R. Kaipa *et al.* Enhanced transmission with a graphene-dielectric microstructure at low-terahertz frequencies. *Phys. Rev. B* **85,** 245407 (2012).
3. A. Fallahi & J. P.-Carrier. Design of tunable biperiodic graphene metasurfaces. *Phys. Rev. B* **86,** 195408 (2012).
4. K. V. Sreekanth, A. De Luca & G. Strangi. Negative refraction in graphene-based hyperbolic metamaterials. *Appl. Phys. Lett.* **103,** 023107 (2013).
5. I. V. Iorsh, I. S. Mukhin, I. V. Shadrivov, P. A. Belov & Y. S. Kivshar. Hyperbolic metamaterials based on multilayer graphene structures. *Phys. Rev. B* **87,** 075416 (2013).
6. M. A. K. Othman, C. Guclu & F. Capolino. Graphene-based tunable hyperbolic metamaterials and enhanced near-field absorption. *Opt. Express* **6,** 7614 (2013).
7. M. A. K. Othman, C. Guclu & F. Capolino. Graphene-dielectric composite metamaterials: evolution from elliptic to hyperbolic wavevector dispersion and the transverse epsilon-nearzero condition. *J. Nanophotonics* **7,** 073089 (2013).
8. Tianrong, Z., Shi, X., Liu, X. & Zi, J. Transfer matrix method for optics in graphene layers. *J. Physics: Conden. Matt.* **25,** 215301 (2013).
9. Y. V. Bludov, N. M. R. Peres & I. M. Vasilevskiy. Unusual reflection of electromagnetic radiation from a stack of graphene layers at oblique incidence. *J. Opt.* **15,** 114004 (2013).
10. Batrakov, K. *et al.* Flexible transparent graphene/polymer multilayers for efficient electromagnetic field absorption. *Sci. Rep.* **4,** 7191 (2014).
11. L. Zhang *et al.* Tunable bulk polaritons of graphene-based hyperbolic metamaterials. *Opt. Express* **22,** 14022 (2014).
12. Y. Xiang *et al.* Critical coupling with graphene-based hyperbolic metamaterials. *Sci. Rep.* **4,** 5483 (2014).
13. A. Andryieuski & A. V. Lavrinenko. Graphene metamaterials based tunable terahertz absorber: effective surface conductivity approach. *Opt. Express* **21,** 9144 (2013).
14. Y. Zhang, Y. Feng, B. Zhu, J. Zhao & T. Jiang. Graphene based tunable metamaterial absorber and polarization modulation in terahertz frequency. *Opt. Express* **22,** 22743 (2014)
15. A. A. Sayem, A. Shahriar, M. R. C. Mahdy & Md. S. Rahman. Control of Reflection through Epsilon near Zero Graphene based Anisotropic Metamaterial Proc. *ICECE* 812–815 (2014).
16. T. Zhang, L. Chen & X. Li. Graphene-based tunable broadband hyperlens for far-field subdiffraction imaging at midinfrared frequencies. *Opt. Express* **21,** 20888 (2013).
17. B. Z. Xu, C.-Q. Gu, Z. Li & Z.-Y. Niu. A novel structure for tunable terahertz absorber based on graphene. *Opt. Express* **21,** 23803 (2013).
18. R. Alaee, M. Farhat, C. Rockstuhl & F. Lederer. A perfect absorber made of a graphene micro-ribbon metamaterial. *Opt. Express* **20,** 28017 (2012).
19. J. Zhang *et al.* Coherent perfect absorption and transparency in a nanostructured graphene film. *Opt. Express* **22,** 12524 (2014).
20. Y. Zou, P. Tassin, T. Koschny & C. M. Soukoulis. Interaction between graphene and metamaterials: split rings vs. wire pairs. *Opt. Express* **20,** 12199 (2012).
21. A. Vakil & N. Engheta. Transformation Optics Using Graphene. *Science* **10,** 332, 1291 (2011).
22. N. Papasimakis *et al.* Graphene in a photonic metamaterial. *Opt. Express* **18,** 8353 (2010).
23. L. Yang, J. Deslippe, C.-H. Park, M. L. Cohen & S. G. Louie. Excitonic Effects on the Optical Response of Graphene and Bilayer Graphene. *Phys. Rev. Lett.* **103,** 186802 (2009).
24. A. Ferreira *et al.* Faraday effect in graphene enclosed in an optical cavity and the equation of motion method for the study of magneto-optical transport in solids. *Phys. Rev. B* **84,** 235410 (2011).
25. R. R. Nair *et al.* Fine Structure Constant Defines Visual Transparency of Graphene. *Science* **320,** 1308 (2008).
26. I. Crassee *et al.* Giant Faraday rotation in single- and multilayer graphene. *Nat. Phys.* **7,** 48 (2011).
27. Z. Sun *et al.* Graphene Mode-Locked Ultrafast Laser. *ACS Nano* **4,** 803 (2010).
28. S. Bae *et al.* Roll-to-roll production of 30-inch graphene films for transparent electrodes. *Nat. Nano.* **5,** 574 (2010).
29. G. Jo *et al.* Large-scale patterned multi-layer graphene films as transparent conducting electrodes for GaN light-emitting diodes. *Nanotechnology* **21,** 175201 (2010).
30. F. Xia, T. Mueller, Y. Lin, A. Valdes-Garcia & P. Avouris. Ultrafast graphene photodetector. *Nat. Nano.* **4,** 839 (2009).
31. T. Stauber & G. Gmez-Santos. Plasmons and near-field amplification in double-layer graphene. *Phys. Rev. B* **85,** 075410 (2012).
32. V. C. Nguyen, L. Chen & K. Halterman. Total Transmission and Total Reflection by Zero Index Metamaterials with Defects. *Phys. Rev. Lett.* **105,** 233908 (2010).
33. M. Silveirinha & N. Engheta. Tunneling of Electromagnetic Energy through Subwavelength Channels and Bends using $\varepsilon$- Near-Zero Materials. *Phys. Rev. Lett.* **97,** 157403 (2006).
34. A. Alu, M. Silveirinha, A. Salandrino & N. Engheta. Epsilon-near- zero metamaterials and electromagnetic sources: Tailoring the radiation phase pattern. *Phys. Rev. B* **75,** 155410 (2007).
35. J. S. Marcos, M. G. Silveirinha & N. Engheta. $\mu$-near-zero supercoupling. *Phys. Rev. B* **91,** 195112 (2015).
36. R. W. Ziolkowski. Propagation in and scattering from a matched metamaterial having a zero index of refraction. *Phys Rev. E* **70,** 046608 (2004).
37. W. Park & J. B. Lee. Mechanically tunable photonic crystal structure. *Appl. Phys. Lett.* **85,** 4845 (2004).
38. C. Kee, J. Kim, H. Park, I. Park & H. Lim. Two-dimensional tunable magnetic photonic crystals. *Phys. Rev. B* **61,** 15523 (2000).
39. A. Figotin, Y. A. Godin & I. Vitebsky. Two-dimensional tunable photonic crystals. *Phys. Rev. B* **57,** 2841 (1998).
40. H. Chen, J. Su, J. Wang & X. Zhao. Optically-controlled high-speed terahertz wave modulator based on nonlinear photonic crystals. *Opt. Express* **19,** 3599 (2011).
41. Merano, M. Fresnel coefficients of a two-dimensional atomic crystal. *Phys. Rev. A* **93,** 013832 (2016).
42. J. Linder & K. Halterman. Graphene-based extremely wide-angle tunable metamaterial absorber. *Sci. Rep.* **6,** 31225 (2016).
43. D. Headland *et al.* Doped polymer for low-loss dielectric material in the terahertz range. *Opt. Mater. Express* **5,** 1373 (2015).
44. John D. Jackson. *Classical Electrodynamics*. Wiley, third edition, August 1998).
45. M. Lobet, B. Majerus, L. Henrard & P. Lambin. Perfect electromagnetic absorption using graphene and epsilon-near-zero metamaterials. *Phys. Rev. B* **93,** 235424 (2016).

46. J. Lötters, W. Olthuis, P. Veltink & P. Bergveld. The mechanical properties of the rubber elastic polymer polydimethylsiloxane for sensor applications. *J. Micromech. Microeng.* **7,** 145 (1997).
47. T. Niu *et al.* Experimental demonstration of reflectarray antennas at terahertz frequencies. *Opt. Express* **21,** 2875 (2013).
48. T. Niu *et al.* Terahertz reflectarray as a polarizing beam splitter. *Opt. Express* **22,** 16148 (2014).
49. S. Walia *et al.* Flexible metasurfaces and metamaterials: A review of materials and fabrication processes at micro-and nanoscales. *Appl. Phys. Rev.* **2,** 011303 (2015).
50. Y.-S. Jin, G.-J. Kim & S.-G. Jeon. Terahertz dielectric properties of polymers. *J. Korean Phys. Soc.* **49,** 513–517 (2006).
51. J. M. Xue *et al.* Scanning tunnelling microscopy and spectroscopy of ultra-flat graphene on hexagonal boron nitride. *Nature Materials* **10,** 282–285 (2011).
52. D. J. Efetov *et al.* Specular interband Andreev reflections at van der Waals interfaces between graphene and $NbSe_2$. *Nature Physics* **12,** 328–332 (2016).
53. X. Huang, Yun Lai, Z. H. Hang, H. Zheng & C. T. Chan. Dirac cones induced by accidental degeneracy in photonic crystals and zero-refractive-index materials *Nature Materials* **10,** 582 (2011).
54. P. Moitra *et al.* Realization of an all-dielectric zero-index optical metamaterial. *Nature Photonics* **7,** 791 (2013).
55. M. Massaouti *et al.* Eutectic epsilon-near-zero metamaterial terahertz waveguides. *Opt. Lett.* **38,** 1140 (2013).
56. K. C. Huang, P. Bienstman, J. D. Joannopoulos, K. A. Nelson & S. Fan. Field Expulsion and Reconfiguration in Polaritonic Photonic Crystals. *Phys. Rev. Lett.* **90,** 196402 (2003).
57. M. Kafesaki, A. A. Basharin, E. N. Economou & C. M. Soukoulis. THz metamaterials made of phonon-polariton materials. *Photon Nanostruct.: Fund. Appl.* **12,** 376 (2014).

### Acknowledgements
J.L. acknowledges support from the Outstanding Academic Fellows programme at NTNU and the Norwegian Research Council Grant No. 216700 and No. 240806. K.H. was supported in part by the NAWC Nanomaterials Core S&T Network and a grant from the Department of Defense High Performance Computing Modernization Program.

### Author Contributions
J.L. and K.H. both performed the analytical and numerical calculations and contributed to the discussion and writing of the manuscript.

### Additional Information
**Competing financial interests:** The authors declare no competing financial interests.

**How to cite this article**: Linder, J. and Halterman, K. Dynamical tuning between nearly perfect reflection, absorption, and transmission of light via graphene/dielectric structures. *Sci. Rep.* **6**, 38141; doi: 10.1038/srep38141 (2016).

**Publisher's note:** Springer Nature remains neutral with regard to jurisdictional claims in published maps and institutional affiliations.